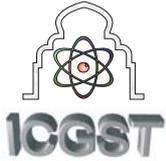

# Cellular Automata based Feedback Mechanism in Strengthening biological Sequence Analysis Approach to Robotic Soccer

P. Kiran Sree [1],    G.V.S. Raju[2],    .S. Viswandha Raju[3] and    N.S.S.S.N Usha Devi[4]

1. Associate Professor, Department of Computer Science, S.R.K  Institute of Technology, Enikepadu,
Vijayawada,India, pkiransree@gmail.com, Mobile: +919959818274.
2. Prof and Head of the Department, Computer Science, Swarnandhra Engineering college,Narasapur.
3. Professor, Department of CSE, Gokaraju Rangaraju Institute of Engineering & Technology
4. Graduate Student of C.S.E,J.N.T.University.

## ABSTRACT

This paper reports on the application of sequence analysis algorithms for agents in robotic soccer and a suitable representation is proposed to achieve this mapping. The objective of this research is to generate novel better in-game strategies with the aim of faster adaptation to the changing environment. A homogeneous non-communicating multi-agent architecture using the representation is presented. To achieve real-time learning during a game, a bucket brigade algorithm is used to reinforce Cellular Automata Based Classifier. A technique for selecting strategies based on sequence analysis is adopted.

## 1. Introduction

Although each domain presents a variety of approaches, from a research perspective the ideal domain embodies as many issues as possible. Robotic soccer is a particularly good domain for studying multi-agent systems. Originated by Alan Mackworth [1], it has been gaining popularity in recent years with several international competitions taking place [2]. Even though robotic soccer is a game, most real-world complexities are maintained. Some of the distinguishing characteristics of the domain include: real-time, noisy with hidden state, collaborative and adversarial goals. It is the ideal test bed for evaluating different machine learning techniques in a direct manner, as it provides for multiple levels of evaluation such as evaluation of low-level behavior e.g. maneuvering on the field as well as evaluation of strategic responses to changing scenarios.

 Several Multi-agent system scenarios are possible:

- Homogeneous, on-communicating
- Homogeneous, communicating
- Heterogeneous, on-communicating
- Heterogeneous, communicating

Each of which have their own set of challenging research issues. In homogeneous, non-communicating multi-agent systems, all of the agents have the same internal structure including goals, domain knowledge, and possible actions. They also have the same procedure for selecting among their actions. The only differences among agents are their sensory inputs and the actual actions they take: they are situated differently in the world. A DNA fragment is usually written as a sequence of letters A, C, T and G – representing the four nucleotides Adenine, Cytosine, Thymine and Guanine. DNA is the master behind all the activities in the cell and is responsible for the synthesis of proteins. In this paper, Section-2 gives an overview of related work. A representation scheme for robotic soccer is presented in Section-3. An architecture depicting a homogeneous non-communicating agent utilizing the above representation is proposed in Section-4. Section-5 discusses experimental results and Section-6 deals with future directions and concludes the paper.

## 2. Related Work

 Balch and Arkin [6] use homogeneous, reactive, non-communicating agents to study formation maintenance in autonomous robots. The robots' goal is to move together in a military formation such as a diamond, column, or wedge. They periodically come across obstacles which prevent one or more of the robots from moving in a straight line. After passing the obstacle, all robots must adjust in order to regain their formation. The actual robot motion is a simple weighted sum of these vectors. Levy and Rosenschein [7] create agents that each act in service of its own goals. They use game theoretic techniques to find equilibrium points and thus to decide how to act. These agents are clearly deliberative, as they search for actions rather than simply retrieving them. There are also several existing systems and techniques that mix reactive and deliberative behaviors. One example is the OASIS system which reasons about when to be reactive and when to follow goal-directed plans [8]. Another example is reactive deliberation [9]. As the name implies, it mixes reactive and deliberative behavior: an agent reasons about which reactive behavior to follow under the constraint that it must choose actions at a rate of 60 Hz. Reactive deliberations was developed on the first robotic soccer platform.

## 3. Sequential representation of robotic soccer

The game of robotic soccer involves a vast amount of data with respect to both player and ball movement. The movement of the ball during the course of a match is tracked to obtain a game sequence. Also, each individual's contribution to the game can be assessed by focusing on player movements during the game, yielding a player sequence. This is instrumental in characterizing a player's contribution to the success or failure of the game. For example, we consider a game played over a period of 60 minutes and represent this game as a sequence of length 60, where each letter represents the player who is in possession of the ball at that minute. Game 1: bcgad-bccc-g-aab—bbaggd. and the Player sequence for player 'a' would be represented as –

Player a: ACTTGC-AG--TTCCCA-...          .

where '-' denotes an idle action or that the ball is in motion. A history of games played can be collected and analyzed using the sequence analysis algorithms employed frequently in bio-informatics.

### 3.1 Algorithm for Finding Tandem Repeats

The concept of finding tandem repeats has been employed here to trace subsequences or regions that frequently repeat in the game and whether such repetition is ideal for the game.

It involves two subtasks
- Finding all possible unique subsequences of a specified length.
- Finding the location and number of tandem repeats.

The input to the first subtask is the starting and ending lengths of the pattern and the result is the creation of a list of unique patterns that range from the specified lower to the upper range. This list becomes the input to the second stage which checks the provided input sequence against these patterns for matches. Table 2 gives the different patterns and occurrences of players.

Table 1: Task Table

| Action | Base |
|---|---|
| Turn towards ball | A |
| Move towards ball | C |
| Kick towards goal | G |
| Pass to team-mate | T |

Table 2: Distinct patterns and their occurrence in each player sequence

| PATTERN | NO. OF OCCURENCES | PLAYER |
|---|---|---|
| CACC | 4 | player-a |
| ACCC | 3 | player-b |
| ACCC | 3 | player-d |
| ACCC | 5 | player-c |
| CCCC | 5 | player-a |
| CCCCB | 2 | player-b |
| CCCAGCC | 2 | player-a |
| CC-G | 3 | player-d |

The agent has just four actuators for physically manipulating the world: turn, dash, kick and catch. The server only executes one of these commands for each player at the end of each simulator cycle. If an agent sends more than one such command during the same cycle, only one is executed non-deterministically. Since the simulator runs asynchronously from the agents, there is no way to keep perfect time with the server's cycle. Therefore any given command could be missed by the server. It is up to the agent to determine whether a given command has been executed by observing the future state of the world.

## 4. Proposed Work

We propose soccer-playing agent architecture with a sequential analysis component, which was strengthened by using CA classifier for providing feed back during the match. We utilize the representation described in the previous sections.

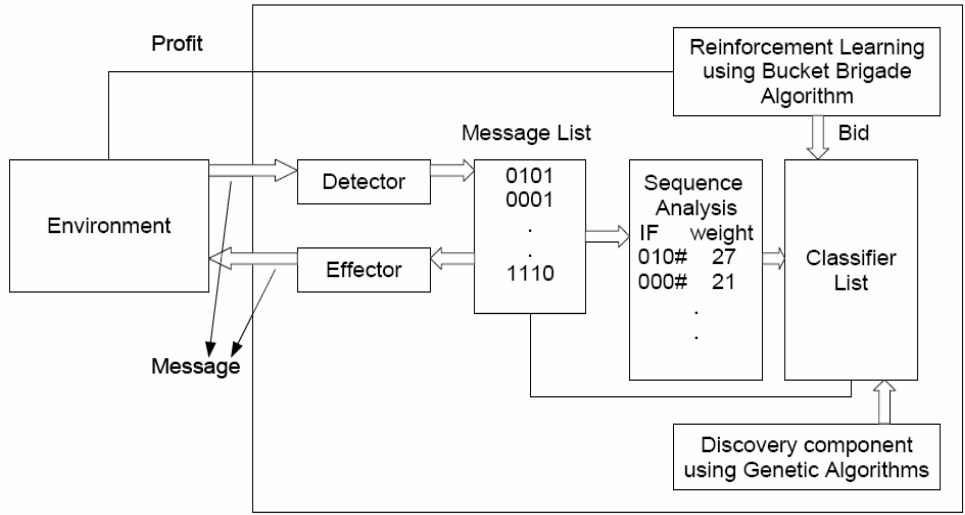

Figure. 1 : Sequence Driven Classifier System Architecture

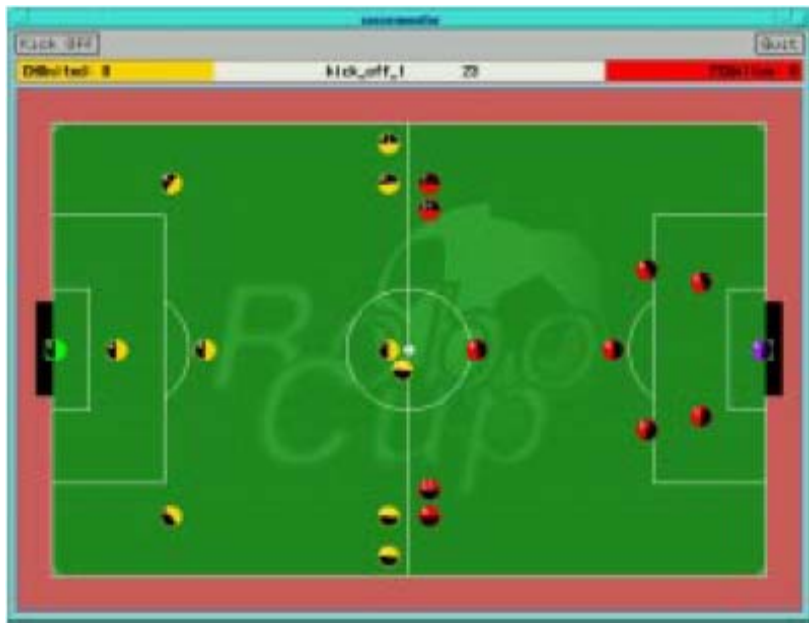

Figure 2: The soccer server display. Each player is represented as a two-halved circle.  The light side is the side towards which the player is facing. All players are facing the ball, which is in the middle of the field. The black bars on the left and right sides of the field are the goals

**Shooting Behavior**
We make a simple strategy for shooting the ball into the goal. To shoot the ball to the goal, it is important that the robot can see both ball and goal. Therefore, the robot must round the ball until the robot can see both ball and goal with the camera toward the ball. Finally, the robot kicks the ball strongly. The concrete procedure of shooting behavior is follows:

1)Find the ball
2)Approach the ball
While approaching the ball
if the area of the ball > 20 then stop

3)Round the ball
d ←the direction of the goal
switch(d) **right:** clockwise round the ball(AGGGT)
with the camera toward the ball(ACCCT)
**left**: counterclockwise round the ball(AAACT)
with the camera toward the ball(TTTAC)
if the robot can see both ball and goal then stop
4)Turn the body of the robot towards the ball(ATACT)
5) Use the CA Feed Back
6)Kick the ball strongly(AATAA)

| Syntax | Argument Meaning | Type | Range | Execute When | Frequency Limit |
|--------|------------------|------|-------|--------------|-----------------|
| say(x) | message to be broadcast | ascii text | $0 \leqslant$ character $\leqslant 512$ | Instantly | Teammates only hear 1 every 2 cycles |
| turn(x) | angle to turn | float | $-180 \leqslant x \leqslant 180$ | End of Cycle | 1 per Cycle |
| dash(x) | power to dash | float | $-30 \leqslant x \leqslant 100$ | End of Cycle | 1 per Cycle |
| kick(x,y) | power to kick, angle to kick | float / float | $0 \leqslant x \leqslant 100$ / $-180 \leqslant y \leqslant 180$ | End of Cycle | 1 per Cycle |
| sense_body() | | | | Instantly | 3 per Cycle |
| change_view( x,y) | view quality / view width | discrete / discrete | high/low / narrow/normal | Instantly | 1 per Cycle |

Table 3: The Soccer server agent's commands

## 4.1. Cellular Automata (CA) and Fuzzy Cellular Automata (FCA)

A CA [4], [5], [6], consists of a number of cells organized in the form of a lattice. It evolves in discrete space and time. The next state of a cell depends on its own state and the states of its neighboring cells. In a 3-neighborhood dependency, the next state $q_i(t+1)$ of a cell is assumed to be dependent only on itself and on its two neighbors (left and right), and is denoted as

$$q_i(t+1) = f(q_{i-1}(t), q_i(t), q_{i+1}(t)) \qquad (1)$$

where $q_i(t)$ represents the state of the $i^{th}$ cell at $t^{th}$ instant of time, $f$ is the next state function and referred to as the rule of the automata. The decimal equivalent of the next state function, as introduced by Wolfram, is the rule number of the CA cell. In a 2-state 3-neighborhood CA, there are total 256 distinct next state functions.

### 4.1.1 FCA Fundamentals

FCA [2], [6] is a linear array of cells which evolves in time. Each cell of the array assumes a state $q_i$, a rational value in the interval [0, 1] (fuzzy states) and changes its state according to a local evolution function on its own state and the states of its two neighbors. The degree to which a cell is in fuzzy states 1 and 0 can be calculated with the membership functions. This gives more accuracy in finding the coding regions. In a FCA, the conventional Boolean functions are AND , OR, NOT.

### 4.1.2 Dependency Matrix for FCA

Rule defined in equation 1 should be represented as a local transition function of FCA cell. That rules are converted into matrix form for easier representation of chromosomes [16].

**Example 1**: A 4-cell null boundary hybrid FCA with the following rule

< 238, 254, 238, 252 > (that is, < $(q_i+q_{i+1})$, $(q_{i-1}+q_i+q_{i+1})$, $(q_i + q_{i+1})$, $(q_{i-1} + q_i)$ >) applied from left to right, may be characterized by the following dependency matrix While moving from one state to other, the dependency matrix indicates on which neighboring cells the state should depend. So cell 254 depends on its state, left neighbor, and right neighbor figure (3). Now we represented the transition function in the form of matrix. In the case of complement FMACA we use another vector for representation of chromosome.

| Non-complemented Rules | | Complemented Rules | |
|------------------------|------------|--------------------|----|
| Rule | Next State | Rule | Next State |
| 0 | 0 | 255 | 1 |
| 170 | $q_{i+1}$ | 85 | $\overline{q_{i+1}}$ |
| 204 | $q_i$ | 51 | $\overline{q_i}$ |
| 238 | $q_i + q_{i+1}$ | 17 | $\overline{q_i + q_{i+1}}$ |
| 240 | $q_{i-1}$ | 15 | $\overline{q_{i-1}}$ |
| 250 | $q_{i-1} + q_{i+1}$ | 5 | $\overline{q_{i-1} + q_{i+1}}$ |
| 252 | $q_{i-1} + q_i$ | 3 | $\overline{q_{i-1} + q_i}$ |
| 254 | $q_{i-1} + q_i + q_{i+1}$ | 1 | $\overline{q_{i-1} + q_i + q_{i+1}}$ |

Table 4.  CA Fuzzy Rules

$$T = \begin{bmatrix} 1 & 1 & 0 & 0 \\ 1 & 1 & 1 & 0 \\ 0 & 0 & 1 & 1 \\ 0 & 0 & 1 & 1 \end{bmatrix}$$

Fig3:  Matrix Representation

### 4.1.4 FMACA Based Tree-Structured Classifier

Cellular Automata choose the rules carefully after trying many other possibilities, some of which caused the cells to die too fast and others which caused too many cells to be born. Life balances these tendencies, making it hard to tell whether a pattern will die out completely, form a stable population, or grow forever. Life is just one example of a **cellular automaton**, which is any system in which rules are applied to cells and their neighbors in a regular grid.

A good way to get started in Life is to try out different patterns and see what happens. Even completely random starting patterns rapidly turn into Life objects recognizable to anyone with a little experience. In this section, we follow a simple-looking pattern called the R-pantomime. It starts out with just five cells, but gets complicated very fast. We can see many of the early discoveries in Life just by running this one pattern in the applet  Like decision tree classifiers, FMACA based tree structured classifier uses the distinct k-means algorithm recursively partitions the training set to get nodes (attractors of a FMACA) belonging to a single class. Each node (attractor basin) of the tree is either a leaf indicating a class; or a decision (intermediate) node which specifies a test on a single FMACA, according to equations 1,2. Suppose, we want to design a FMACA based pattern classifier to classify a training set $S = \{S1, S2, \cdot, SK\}$ into $K$ classes. First, a FMACA with $k$-attractor basins is Once we formulated the transition function, we can move form one state to other. For the example 1 if initial state is P (0) = (0.80, 0.20, 0.20, 0.00) then the next states will be

P (1) = (1.00 1.00, 0.20, 0.20),
P (2) = (1.00 1.00, 0.40, 0.40),
P (3) = (1.00 1.00, 0.80, 0.80),
P (4) = (1.00 1.00, 1.00, 1.00).

We just introduced the concept of Cellular Automata in this section. CA is a linear array of cells which evolves in time. Each cell of the array assumes a state $qi$, a rational value in the interval [0, 1] (fuzzy states) and changes its state according to a local evolution function on its own state and the states of its two neighbors. The degree to which a cell is in fuzzy states 1 and 0 can be calculated with the membership functions. This gives more accuracy in finding the coding regions. In a FCA, the conventional Boolean functions are AND , OR, NOT. Figure 1,2 shows the architecture of the proposed sequence-driven classifier system. This system differs from standard classifier systems **[5]** in two main ways generated. The training set $S$ is then distributed into $k$ attractor basins (nodes). Let, $S'$ be the set of elements in an attractor basin. If $S'$ belongs to only one class, then label that attractor basin for that class. Otherwise, this process is repeated recursively for each attractor basin (node) until all the examples in an attractor basin belong to one class. Tree construction is reported in [7]. The above discussions have been formalized in the following algorithm. We are using genetic algorithm classify the training set.

***Algorithm 1:* FMACA Tree Building (using distinct K means algorithms)**
Input : Training set $S = \{S1, S2, \cdot\cdot, SK\}$
 Output: FMACA Tree.
**Partition**($S$, $K$)
Step 1: Generate a FMACA with $k$ number of attractor basins.
Step 2: Distribute $S$ into $k$ attractor basins (nodes).
Step 3: Evaluate the distribution of examples in each attractor basin (node).
Step 4: If all the examples (S') of an attractor basin (node) belong to only one class, then label the attractor basin (leaf node) for that class.
Step 5: If examples (S') of an attractor basin belong to $K'$ number of classes, then **Partition** (S', $K'$).
Step 6: Stop.

First, the proposed system adds an event-analysis section and prepares a record of past game sequences/player sequences that are analyzed using sequence analysis. Second, the classifier discovery section using genetic algorithms targets only actions while the conditions are generated using information provided by sequence analysis. The systems also provides for different types of reward – a large reward for winning matches and smaller rewards that can be obtained from succeeding in a single play, such as passing the ball

## 5. Experimental Results
The Sequence Analysis algorithms were tested on player and game sequences of varying lengths and distinct sub sequences yielded the following results from which various inferences for improving the next match were made.

Client 1 sends movement commands after every perception it receives. This strategy works out fine in cycle t-1; but in cycle t it misses the opportunity to act because it receives no perceptions; and in cycle t+1 it sends two movement commands, only one of which will be executed. Client 2, on the other hand, successfully sends one movement command every cycle. Note that in cycle t it must act with no new perceptual information, while in cycle t+1, it receives two perceptions prior to acting and one afterwards. Ideally, it would act after receiving and taking into account all three perceptions. However, it does not know precisely when the simulator cycle will change internally in the soccer server and it cannot know ahead of time when it will receive perceptions. Thus, in exchange for the ability to act every simulator cycle, it sometimes acts with less than the maximal amount of information about the world. However, as each simulator cycle represents only a short amount of real time (*simulator_step* msec), the world does not change very much from cycle to cycle, and the client can act accurately even if it takes some of its perceptions into account only before its subsequent action.

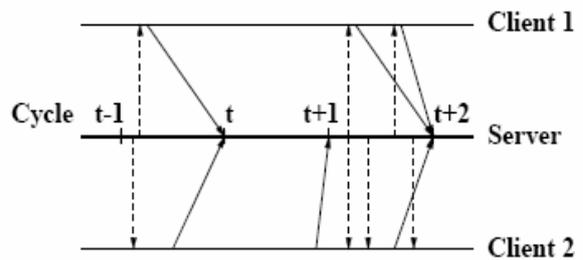

Figure 4: A sample period of the server-client interface over the course of 3 simulator cycles at times t-1, t, and t+1. The thick central horizontal line represents the real time as kept by the server. The top and bottom horizontal lines represent the time-lines of two separate clients. The dashed arrows from the server towards a client represent perceptions for that client. The solid arrows from a client towards the server represent movement commands sent by that client. These arrows end at the point in time at which the server executes the movement commands, namely the end of the simulator cycle during which they are sent. Asynchronous sensing and acting, especially when the sensing can happen at unpredictable intervals, is a very challenging paradigm for agents to handle. Agents must balance the need to act regularly

and as quickly as possible with the need to gather information about the environment. Along with asynchronous sensing and action, the soccer server captures several other real-world complexities, as will become evident throughout the remainder of this section.

## 5.1 Entropy

Entropy is the measure of randomness of a system. The maximum entropy (close to 1) of a system signifies chaotic behavior, whereas low entropy indicates ordered behavior. In case of complex system, mean entropy is close to the critical value 0·84 with high variance. To measure the entropy, we select a moving window of 10 time steps ($w$=10). The CA rule vector is operated for 10000 time steps from a random initial state. The mean entropy and the standard deviation from the mean have been computed. For each CA the procedure is repeated for 15 times with different random initial states. The high entropy with high standard deviation indicates that the evolved CA performing pattern recognition task are at the edge of chaos.

## 5.2 Mutual Information

Mutual information measures the correlation between patterns generated at a fixed time interval. If a pattern $P1$ is the copy of $P2$, then mutual information between $P1$ and $P2$ is 1. The mutual information between two statistically independent patterns is 0. Both the ordered and chaotic CA rules do not create highly correlated patterns and in effect generate pattern set with low mutual information. On the other hand, the complex CA rules create highly correlated structures producing maximum mutual information. To measure the mutual information of the patterns generated by CA, we follow the method proposed. These results were translated into strength values and directly modified within the LCS. In order to justify that this proposed architecture is indeed faster at learning targeted behavior at a much faster pace, "Clean Slate" experimental conditions were maintained. The Classifier rule set initially is completely randomly generated – to simulate a blank knowledge base. The rule set is then generated prior to the first iteration and the LCS was run for up to 200,000 iterations. Initialization parameters for the Genetic Algorithm specify the algorithm to run every 4000 iterations – this interval is necessary so that the strengths of the rule sets are given time to stabilize after credit apportionment. The entropy and mutual information of the CA in successive generations of GA are reported in Figure 5,6 ,7,8

For four different CA size ($n$= 10, 15, 20, 30). For each of the cases, the values of entropy and mutual information reach their steady state once the AIS FMACA for a given pattern set gets evolved. For understanding the motion, the initial population (IP) is randomly generated. All these figures points to the fact that as the CA evolve towards the desired goal of maximum pattern recognizing capability, the entropy values fluctuate in the intermediate generations, but saturate to a particular value (close to the critical value 0·84 [245]) when fit rule is obtained. Simultaneously, the values of mutual information fluctuate at the intermediate points prior to reaching maximum value that remains stable in subsequent generations. All these figures indicate that the CA move from chaotic region to the edge of chaos to perform complex computation associated with pattern recognition. Table 3 shows the application of the algorithm for a single game which when applied to hundreds of games, useful inferences could be made. The patterns thus identified through the application of tandem repeats algorithm served to identify the risk of threats and goals thereby enabling dynamic decision making in future games with reference to past history of games played. The goal sequences identified were xxCCT, where x represents any of the 4 characters. The following is a table showing the sequences that are likely to occur in the case of a goal or threat and their percentage of occurrences. The goal sequences were identified by application of the algorithm for finding tandem repeats. These sequences were in turn checked using the algorithm for Trans-membrane region to find the percentage of occurrences of these patterns in multiple numbers of games.

Table 3: Percentage of occurrence of Goal and Threat sequences

| SEQUENCES | 95% | 75% | 50% | < 50% |
|---|---|---|---|---|
| Goal (xxCCT) | TCCCT | CACCT | CxCCT | CCAT |
| Threat | CTCCC | CCACC | CCxCC | GCAC |

The Learning classifier system was first tested by first evolving completely random classifier rules and then placed in the environment to gage the learning rate. Figure 2 shows the proportion of correct actions proposed by the learning classifier system versus the time epochs taken. Subsequent experiments were conducted by constructing a history of 10,000 games – and the performance of the classifier system was observed. Figure 7. shows the performance of LCS, an accuracy rate of nearly 75% was obtained after merely 100,000 iterations. This is observed to be far less when compared to other agent learning mechanisms based on reinforcement learning. Given a good balance between desired accuracy and training time, this technique will yield a good "knowledge base" of conditions and actions upon which the agent will base all its responses.

## 6 .Conclusion

The advantage of adopting sequence analysis for evaluating strategy is that the analysis is capable of distinguishing individual strategy of an agent as well the overall strategic play of the team as such. This approach could easily be adapted towards developing squad-based tactics of team behavior evolution, by evolving specific strategies for specific groups - for example, creating a group of defenders whose overall goal is to defend their goal, and a squad of offensive players whose only aim is to target the opposing team's goal. Furthermore, use of advanced genetic operations during the discovery stages could enable the classifier system to discover better actions within a smaller time-bound and improve the real-time response of the system. A representation scheme for robotic soccer in the form of biological sequence has been presented. Two new algorithms to find repeating patterns and ideal game strategies similar to biological problems have been developed and tried on the sequences generated. A number of sequences has been generated from various games played and the algorithms proved effective in analysing various repeating patterns before a goal and finding faulty moves. The entire system was improved by using the feed back generated by CA Classifier.

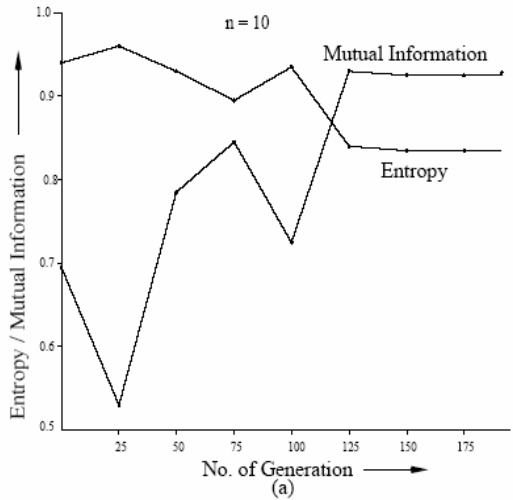

Figure 5: Entropy & Mutation information for n=10

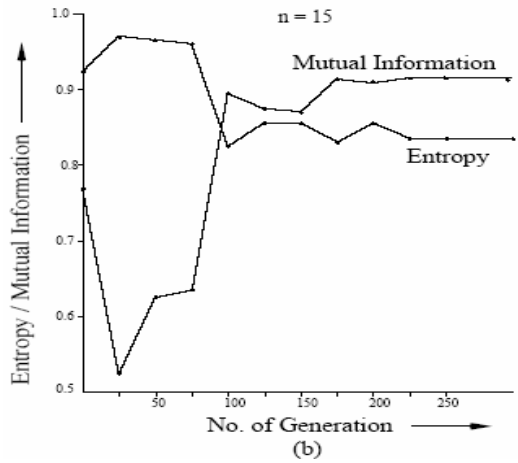

Figure 6: Entropy & Mutation information for n=15

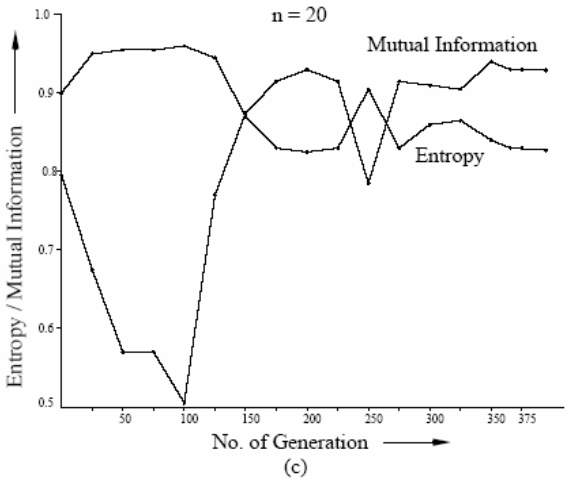

Figure 7: Entropy & Mutation information for n=20

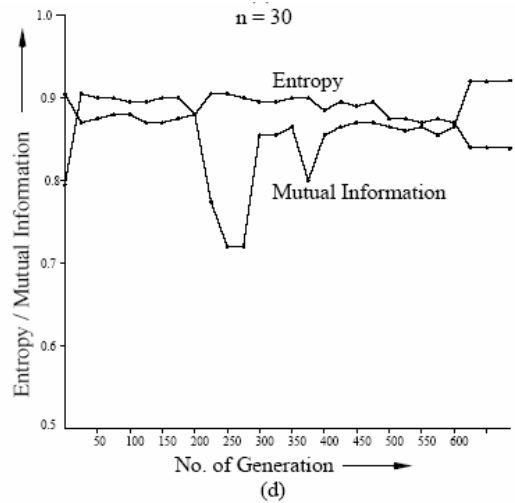

Figure 8: Entropy & Mutation information for n=30

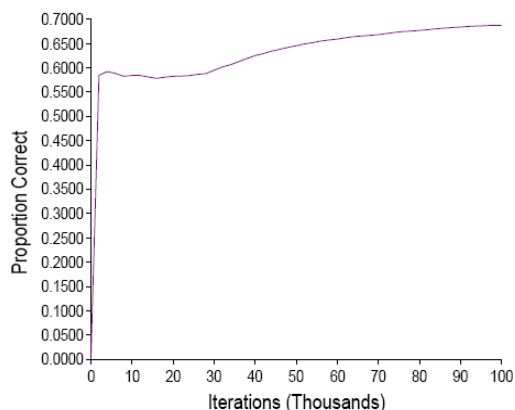

Figure 9. Graph (Proportion Correct VS Iterations)

# 7. References

[1] Michael Sahota, Alan K. Mackworth, Rod A. Barman, and Stewart J. Kingdon , "Real-time control of soccer-playing robots using off-board vision: the dynamite testbed", *IEEE International Conference on Systems, Man, and Cybernetics*, pages 1995 ,3690-3663.

[2] Hiroaki Kitano, Yasuo Kuniyoshi, Itsuki Noda, Minoru Asada, Hitoshi Matsubara, and Ei-Ichi Osawa, " Robocup: A challenge problem for ai ", *AI Magazine*, Volume 18(Issue 1), 1997,pp. 73-85.

[3] Needleman, S.B., Wunsch C.D., "A general method applicable to the search for similarities in the amino acid sequences of two proteins", *Journal of Molecular Biology* Volume 48, 2000, pp. 443-453.

[4] Smith T.F, Waterman M.S., "Identification of common molecular subsequences" *Journal of Molecular Biology*, Volume 14, 1998,pp.95-197.

[5] Goldberg, D.E. , " Genetic algorithms in Search ,Optimization and Machine Learning ", Addison-Wesley, 1989,Reading.

[6] Tucker Balch and Ronald C. Arkin. , " Motor schema-based formation control for multi agent robot teams ", *Proceedings of the First International Conference on Multi-Agent Systems ICMAS-95*, 1995,pp. 10-16.

[7] Ran Levy and Jeffrey S. Rosenschein , "A game theoretic approach to the pursuit problem ", *Working Papers of the 11th International Workshop on Distributed Artificial Intelligence*, 1992,pp.195-21.

[8] Anand S. Rao and Michael P. George , " BDI agents: From theory to practice.", *Proceedings of the First International Conference on Multi-Agent Systems ICMAS-95*, 1995, pp. 31

[9] Michael K. Sahota. , "Reactive deliberation: An architecture for real-time intelligent control in dynamic environments", *Proceedings of the Twelfth National Conference on Artificial Intelligence*, pages 1303-1308,1994, pp. 184, 185, 202.

[10] P.Kiran Sree, I .Ramesh Babu ,"Identification of Protein Coding Regions in Genomic DNA Using Unsupervised FMACA Based Pattern Classifier" in International Journal of Computer Science & Network Security with ISSN: 1738-7906 Volume Number: Vol.8, No.1,2008.

[11] P.Kiran Sree, R.Ramachandran, "Identification of Protein Coding Regions in Genomic DNA Using Supervised Fuzzy Cellular Automata".in International journal of Advances in Computer Science and Engineering, with ISSN: 0973-6999, Vol: 1,2008.

[12] Eric E. Snyder ,Gary D. Stormo, " Identification of Protein Coding Regions In Genomic DNA".ICCS Transactions 2002.

[13] E E Snyder and G D Stormo,"Identification of coding regions in genomic DNA sequences: an application of dynamic programming and neural networks " Nucleic Acids Res. 1993 February 11; 21(3): 607–613.

[14] P. Flocchini, F. Geurts, A. Mingarelli, and N. Santoro (2000),"Convergence and Aperiodicity in Fuzzy Cellular Automata: Revisiting Rule 90,"Physica D.

[15] P. Maji and P. P. Chaudhuri (2004),"FMACA: A Fuzzy Cellular Automata Based Pattern Classifier," Proceedings of 9th International Conference on Database Systems , Korea, pp. 494–505, 2004.

[16] C.G. Langton (2000), "Self-reproduction in cellular automata," Physica D, vol.10, pp.135–144.

[17] T. Toffoli(1998), "Reversible computing," in Automata, Languages and Programming, ed. J.W. De Bakker and J. Van Leeuwen, pp.632–644.

[18] G. Vichniac(1994), "Simulating physics with cellular automata," Physica D,vol.10, pp.96–115.

[19] S. Chattopadhyay, S. Adhikari, S. Sengupta, and M. Pal (2000), "Highly regular, modular, and cascadable design of cellular automata-based pattern classifier," IEEE Trans. Very Large Scale Integr. Syst., vol.8, no.6,

[20] J. Fickett (1982), "Recognition of protein coding regions in dna sequences,"Nucleic Acids Res., vol. 10, pp. 5303–5318.

[21] B. E. Blaisdell(1983), "A prevalent persistent global non randomness that distinguishes coding and non-coding eukaryotic nuclear dna sequence," J. Molec. Evol., vol. 19, pp. 122–133.

[22] R. Farber, A. Lapedes, and K. Sirotkin(1992), "Determination of eukaryotic protein coding regions using neural networks and information theory," J. Mol. Biol., vol. 226, pp. 471–479

[23] E. Uberbacher and R. Mural(1991), "Locating protein-coding regions in human dna sequences by a multiple sensor-neural network approach," Proc. Natl. Acad. Sci., USA, vol. 88, pp. 11261–11265.

[24] Aickelin U and Cayzer S (2002): The Danger Theory and Its Application to AIS, Proceedings 1st International Conference on AIS, pp 141-148, Canterbury, UK.

[25] D. Dasgupta, 1999. Artificial Immune Systems and Their Application. Berlin, Germany: Springer-verlag.

[26] Jerne, N. K. (1974), "Towards a Network Theory of the Immune System", Ann. Immunol. (Inst. Pasteur) 125C, pp. 373-389.

## Biography

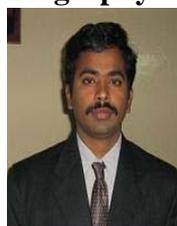

**P.KIRAN SREE** received his B.Tech in Computer Science & Engineering, from J.N.T.U and M.E in Computer Science & Engineering from Anna University. He is pursuing Ph.D in Computer Science from J.N.T.U, Hyderabad. He has published many technical papers; both in international and national Journals& Conferences .His areas of interests include Cellular Automata, Parallel Algorithms, Artificial Intelligence, Compiler Design and Computer Networks. He also wrote books on Analysis of Algorithms, Theory of Computation and Artificial Intelligence. He was the reviewer for many International Journals and IEEE Society Conferences in Artificial Intelligence and Networks. He was also member in many International Technical Committees. He was the Associate Editor for Asian Journal of Scientific Research (ISSN: 1992-1454), Journal of Artificial Intelligence (ISSN: 1994-5450), Information Technology Journal (ISSN: 1812-5646), Journal of Software Engineering (ISSN: 1819-4311), and Research Journal of Information Technology. He was also invited speaker and organizing chairman for special sessions in WSEAS international conferences. He is the member of C.S.I, I.E.T.E, I.S.T.E (India), ICST (Europe) and IAENG (U.S.A). He is now associated with S.R.K Institute of Technology, Vijayawada

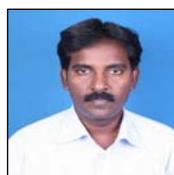

**G.V.S.Raju** is working as a Professor in the Department of C.S.E, Swarnandhra Engineering College, Narasapuram, West Godavari , Dt. He has published good number of papers in international conferences.

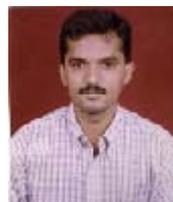

**Dr S. Viswanadha Raju** is working as a Professor in Department of CSE, Gokaraju Rangaraju Engineering College, Hyderabad. He has published good number of papers both in international conferences and journals

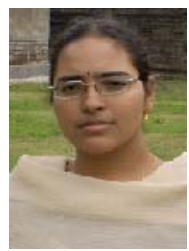

**N.S.S.S.N Usha Devi** was a graduate student of C.S.E from J.N.T.U. She has published 5 research papers in international conferences and two in international journals